\documentclass[10pt, twocolumn, notitlepage, superscriptaddress, prl, longbibliography, colorlinks=true, linkcolor=blue, citecolor=blue]{revtex4-2}

\usepackage{here}
\usepackage{amsmath}         
\usepackage{graphicx}
\usepackage{braket}         
\usepackage{lettrine}
\usepackage{color}
\usepackage{times}
\usepackage{hhline}
\usepackage{tabularx}
\usepackage{calc}
\usepackage{soul}
\usepackage{booktabs}
\usepackage{orcidlink}
\usepackage{makecell}

\usepackage{hyperref}

\begin{document}

\title{Multipolar Anisotropy of Magnonic Orbital Pumping in $d$-Wave Altermagnets}

\author{Myeonil Choi\,\orcidlink{0009-0008-8834-6976}}
\affiliation{Department of Physics, Korea University, Seoul 02841, Korea}
\affiliation{Center for Quantum Dynamics of Angular Momentum, Pohang 37673, Korea}

\author{Guidobeth Sáez\,\orcidlink{0000-0003-1401-2531}}
\affiliation{Peter Grünberg Institut, Forschungszentrum Jülich, 52425 Jülich, Germany}

\author{Daegeun Jo\,\orcidlink{0000-0002-0655-4552}}
\affiliation{Department of Physics and Astronomy, Uppsala University, P.O. Box 516, SE-75120 Uppsala, Sweden}
\affiliation{Wallenberg Initiative Materials Science for Sustainability, Uppsala University, SE-75120 Uppsala, Sweden}

\author{Yuriy Mokrousov\,\orcidlink{0000-0003-1072-2421}}
\affiliation{Peter Grünberg Institut, Forschungszentrum Jülich, 52425 Jülich, Germany}
\affiliation{Institute of Physics, Johannes Gutenberg University Mainz, 55099 Mainz, Germany}

\author{
Hyun-Woo Lee\,\orcidlink{0000-0002-1648-8093}}
\affiliation{Department of Physics, Pohang University of Science and Technology, Pohang 37673, Korea}
\affiliation{Center for Quantum Dynamics of Angular Momentum, Pohang 37673, Korea}

\author{Kyoung-Whan Kim\,\orcidlink{0000-0001-7053-4372}}
\affiliation{Department of Physics, Yonsei University, Seoul 03722, Korea}
\affiliation{Center for Quantum Dynamics of Angular Momentum, Pohang 37673, Korea}

\author{Dongwook Go\,\orcidlink{0000-0001-5740-3829}}
\thanks{Corresponding author: \href{mailto:dongwookgo@korea.ac.kr}{dongwookgo@korea.ac.kr}}
\affiliation{Department of Physics, Korea University, Seoul 02841, Korea}
\affiliation{Center for Quantum Dynamics of Angular Momentum, Pohang 37673, Korea}

\begin{abstract}
We show that antiferromagnetic magnons in $d$-wave altermagnets lead to orbital pumping with distinctive $d$-wave angular dependence on the direction of the N\'eel vector. We find that the anisotropy is generally much larger than that of spin pumping. We elucidate that this difference originates from the direct coupling of the orbital response to the electrical quadrupole moments induced by staggered local crystal field in an altermagnet. By symmetry analysis, we show that the anisotropy does not cancel for the two opposite-handedness antiferromagnetic magnons, suggesting that our prediction can be observed even with thermally excited magnons. We further analyze the handedness dependence for the situation with coherent magnons. These results establish multipolar anisotropy of orbital pumping as a direct dynamical signature of $d$-wave altermagnetic order and highlight the potential of altermagnets as a platform for magnon-driven orbitronics as well as for the interplay between magnons and electrical multipoles.
\end{abstract}

\date{\today}                 
\maketitle

Altermagnets are collinear magnets in which the opposite-spin sublattices
are related by a crystal rotation rather than by translation or inversion.
This symmetry produces a nonrelativistic, momentum-dependent spin splitting
even though the net magnetization is fully compensated
\cite{vsmejkal2022beyond, mazin2022altermagnetism, fender2025altermagnetism}.
Photoemission has since resolved altermagnetic band structures in a growing
set of compounds \cite{krempasky2024altermagnetic, reimers2024direct,
osumi2024observation, lee2024broken, zhu2024observation}. In $d$-wave altermagnets, where
the splitting reverses sign under a $90^\circ$ rotation of momentum, the spin splitting has been observed in the layered metals KV$_2$Se$_2$O and
Rb$_{1-\delta}$V$_2$Te$_2$O \cite{jiang2025metallic, zhang2025crystal}, rutiles such as MnF$_2$ \cite{vsmejkal2022emerging, fan2025high, yuan2020giant} and the
much-debated RuO$_2$ \cite{fedchenko2024observation, hiraishi2024nonmagnetic,
kessler2024absence, liu2024absence} among further candidates. The resulting
phenomena---spin-splitter currents \cite{hallberg2025visualization,
sarkar2026extrinsic, gonzalez2021efficient}, tunneling magnetoresistance
\cite{gonzalez2024anisotropic, noh2025tunneling}, current-induced torques
\cite{zhang2025electrical, han2026deterministic, bai2022observation, karube2022observation}, and nonlinear optical
responses \cite{yang2025nonlinear}---comes together with the terahertz
dynamics and vanishing stray fields of an antiferromagnet. Yet most existing probes, however, access altermagnetism through spin splitting or its spin-dependent transport it produces. A complete understanding of altermagnetism requires consideration of the underlying crystalline and multipolar order, which couples to the orbital degree of freedom. A dynamical response that directly reflects this order is therefore highly desirable to fully exploit the potential of altermagnets.

The defining symmetry of an altermagnet, however, is not encoded in the spin order alone. Because the opposite-spin sublattices are related by crystal rotations, the N\'eel order is intertwined with a sublattice-dependent orbital environment. In $d$-wave altermagnets, this structure may be viewed as a staggered electric quadrupolar order of the local charge coupled to the collinear magnetic dipolar order \cite{McClarty2024, bhowal2024ferroically, lh4w-byk7}. Such multipolar ordering underlies recently proposed phenomena in altermagnets, such as magnetoelectric coupling \cite{oike2024nonlinear, verbeek2024nonrelativistic}, octupole Hall effect \cite{ko2025magnetic, baek2025magnetic}, current-induced torques \cite{han2025harnessing} and deterministic control of the N\'eel order~\cite{han2026deterministic}. These developments show that orbital and multipolar degrees of freedom are not merely secondary consequences of altermagnetic spin splitting, but constitute an intrinsic component of the altermagnetic order. Altermagnets therefore provide a natural platform for {\it orbitronics} \cite{go2021orbitronics}, in which orbital angular momentum (OAM) can be generated \cite{go2020orbital, go2017toward, park2011orbital, yoda2018orbital, salemi2019orbitally, lee2021orbital, kim2021nontrivial, sala2022giant, lee2021efficient, hayashi2023observation}, transported \cite{go2018intrinsic, bernevig2005orbitronics, tanaka2008intrinsic, sala2022giant}, and manipulated through its coupling to magnetic and crystalline order. Despite this intrinsic synergy, the nonequilibrium orbital response of altermagnets to collective magnetic excitations remains largely unexplored.

Orbital pumping \cite{hayashi2024observation, go2025orbital, han2025orbital, wang2025orbital} provides a natural route to access the coupling between magnetic order and the OAM of electrons in dynamic regime. Magnonic pumping converts the collective motion of local moments into a nonequilibrium angular-momentum response of itinerant electrons, thereby providing a direct probe of electron--magnon interactions. In conventional magnets, the OAM generated by magnetization dynamics is commonly regarded as a secondary response accompanying spin pumping through spin-orbit coupling (SOC)~\cite{go2025orbital, hayashi2024observation}. In altermagnets, by contrast, antiferromagnetic magnons modulate a magnetic order that is intrinsically intertwined with the staggered quadrupolar crystal-field and orbital configuration of electrons~\cite{McClarty2024, bhowal2024ferroically, jo2025weak}. 

\begin{figure}[t!]
\includegraphics[angle=0, width=0.47\textwidth]{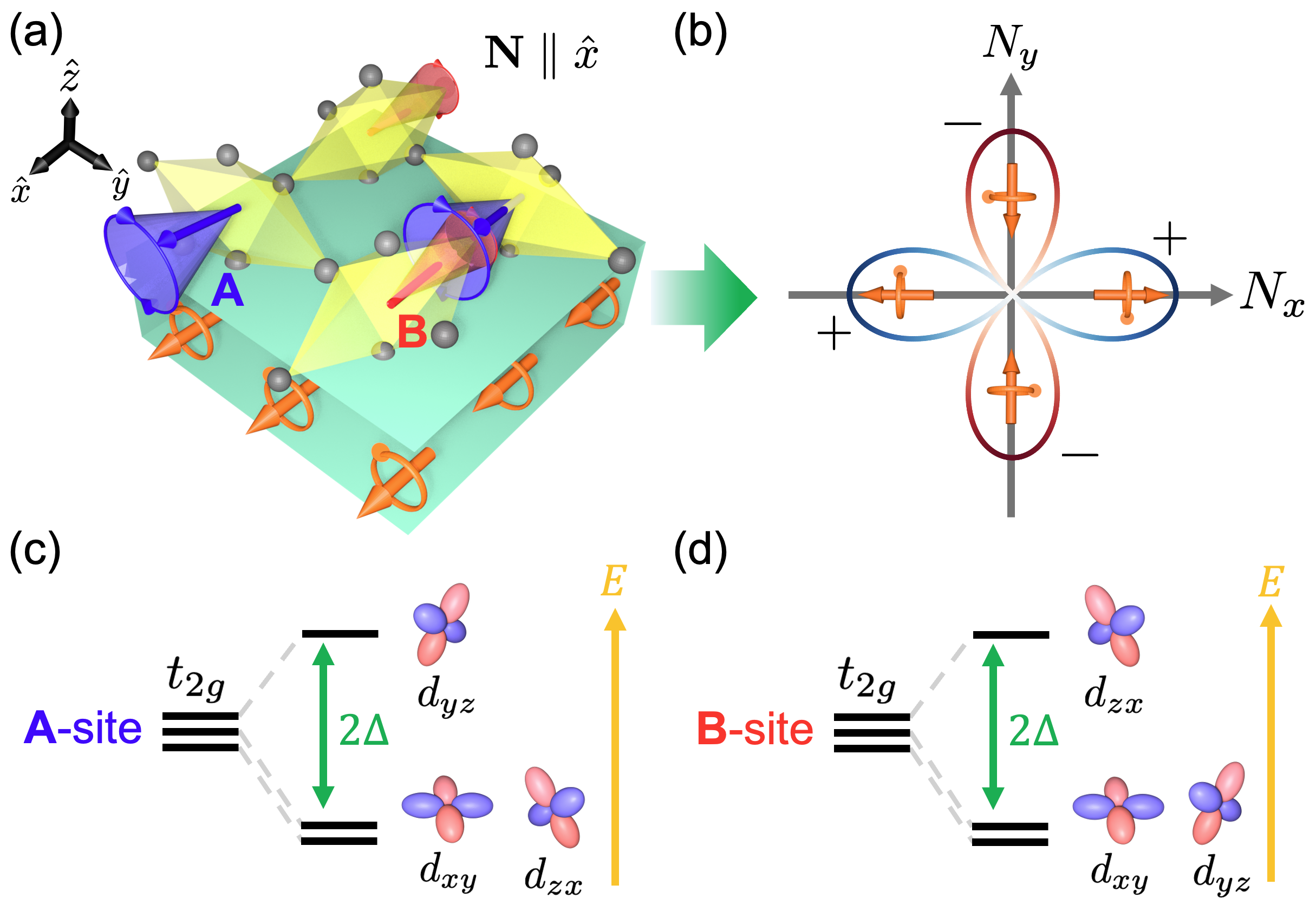}
\caption{
\label{fig:schematics}
\label{Fig1}
Schematic illustration of anisotropic orbital pumping in a $d$-wave altermagnet.
(a) An antiferromagnetic magnon drives the dynamics of local moments on sublattices A (blue) and B (red) with different precession cone angles and staggered crystal field (yellow octahedra), generating the OAM (orange).
(b) The resulting OAM accumulation along the N\'eel-vector direction contains a pronounced $d$-wave angular modulation. This originates from the mutually rotated crystal fields, responsible for (c,d) alternating $t_{2g}$ orbital splittings on sublattices A and B, respectively.
}
\end{figure}

In this Letter, we show that orbital pumping in a $d$-wave altermagnet exhibits substantial anisotropy, following the symmetry of underlying spin-splitting. As illustrated in Fig.~\ref{Fig1}(a), an antiferromagnetic magnon drives the local moments on sublattices A and B with different precession cone angles. The resulting OAM accumulation exhibits a pronounced angular dependence with respect to the N\'eel-vector orientation, as schematically shown in Fig.~\ref{Fig1}(b). This response originates from the mutually rotated local crystal fields on the two sublattices and the resulting electric quadrupolar order, which produce distinct $t_{2g}$ orbital configurations on sites A and B [Figs.~\ref{Fig1}(c) and \ref{Fig1}(d)]. Orbital pumping can therefore reveal how antiferromagnetic magnons couple to the orbital character of electronic states, through alternating single-ion magnetic anisotropy. This suggests that measurable orbital pumping serves as a  dynamical fingerprint of altermagnetism--entangled electrical quadrupole and magnetic dipole--compared to spin-dependent responses. Our work establishes the connection among antiferromagnetic magnons, multipolar order, and the OAM dynamics in altermagnets.

\begin{figure*}[t!]
\includegraphics[angle=0, width=1\textwidth]{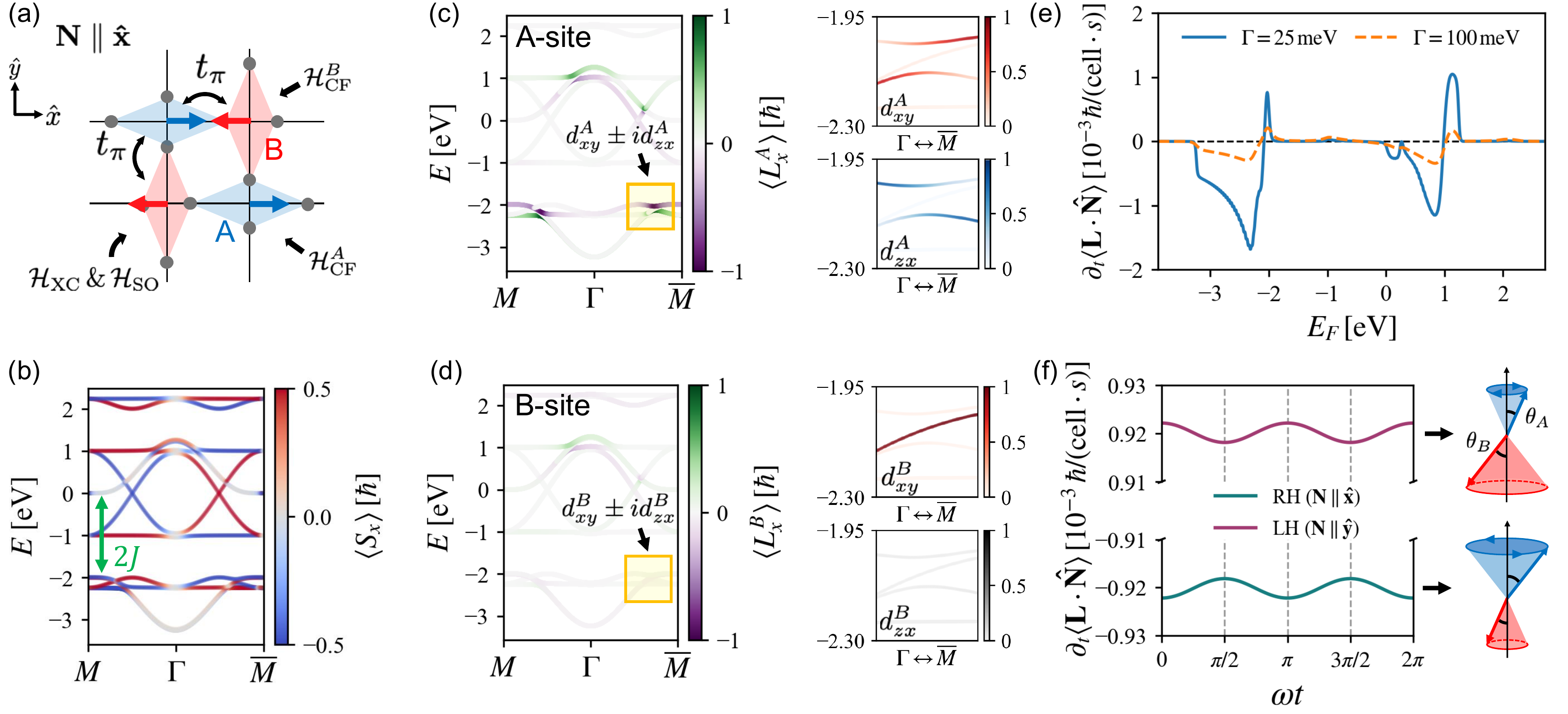}
\caption{
\label{fig:band_structure}
(a) Tight-binding model of a $d$-wave altermagnet with hopping $t_\pi$, SOC $\mathcal{H}_\mathrm{SO}$, exchange coupling $\mathcal{H}_\mathrm{XC}$ with an $x$-aligned Néel vector $\hat{\mathbf{N}}$, and alternating crystal fields, $\mathcal{H}_\mathrm{CF}^\mathrm{A}$ and $\mathcal{H}_\mathrm{CF}^\mathrm{B}$ on sublattices A (blue) and B (red), respectively.
(b) Band structure resolved by $\langle S_x\rangle$; the exchange splitting $J$ is highlighted in green.
(c,d) Band structures resolved by the sublattice OAM on A and B. Insets show the corresponding orbital characters and their weights in the yellow-boxed region.
(e) Fermi-energy dependence of the net OAM pumping along the Néel vector for $\Gamma=25$ meV (blue solid) and $100$ meV (orange dashed).
(f) Time evolution over one precession period for the RH mode with an $x$-aligned Néel vector (green solid) and the LH mode with a $y$-aligned Néel vector (purple dashed).
}
\end{figure*}

To investigate the microscopic origin of orbital pumping in a $d$-wave altermagnet, we consider the $t_{2g}$-orbital tight-binding model illustrated in Fig.~\ref{fig:band_structure}(a) (see Supplemental Material \cite{supp} for details). In the sublattice basis, the Hamiltonian is written as
\begin{equation}
\mathcal{H}_\mathrm{tot}
=
\begin{pmatrix}
\mathcal{H}_{\mathrm{CF}}^\mathrm{A}
+
\mathcal{H}_{\mathrm{XC}}^\mathrm{A}
+
\mathcal{H}_{\mathrm{SO}}
&
h_{\mathrm{hop}}
\\
h_{\mathrm{hop}}^{\dagger}
&
\mathcal{H}_{\mathrm{CF}}^\mathrm{B}
+
\mathcal{H}_{\mathrm{XC}}^\mathrm{B}
+
\mathcal{H}_{\mathrm{SO}}
\end{pmatrix},
\label{eq:Hamiltonian}
\end{equation}
where 
\begin{eqnarray}
h_\mathrm{hop}
&=&
t_\pi \ \mathrm{diag}
\Big[
\cos (k_x/\sqrt{2}a) + \cos (k_y/\sqrt{2}a),\\
\nonumber 
& & \ \ \ \ \ \ \ \ \ \ \ \ \ \ \ \cos (k_y/\sqrt{2}a), \ \  \cos (k_x/\sqrt{2}a)
\Big],
\end{eqnarray}
describes nearest-neighbor hoppings between sublattices A and B, considering only the $\pi$-type hopping $t_\pi$ among the same orbitals, and $a$ is the lattice constant. Importantly, 
\begin{subequations}
\begin{eqnarray}
\mathcal{H}_\mathrm{CF}^\mathrm{A} &=& \Delta \mathrm{diag}[-1,\ +1,\ -1],
\\
\mathcal{H}_\mathrm{CF}^\mathrm{B} &=& \Delta \mathrm{diag}[-1,\ -1,\ +1]
\end{eqnarray}
\end{subequations}
are alternating local crystal-field splittings by $2\Delta$ on sites A and B, respectively, consistent with Fig.~\ref{fig:schematics}(c,d), in $d_{xy}$, $d_{yz}$, and $d_{zx}$-derived Bloch basis \cite{jo2025weak}. These therefore describe mutually rotated local orbital environments, which distinguish the $d$-wave altermagnet from a conventional collinear antiferromagnet. The exchange interaction between the conduction electron's spin $\mathbf{S}$ and the local moments $\hat{\mathbf{m}}^\mathrm{A(B)}$ on site A(B) is $\mathcal{H}_{\mathrm{XC}}^\mathrm{A(B)} = J \hat{\mathbf{m}}^\mathrm{A(B)}\cdot\mathbf{S}$, separately for sites A and B. The SOC is equally included on both sites A and B, $\mathcal{H}_\mathrm{SO}=\lambda \mathbf{L}\cdot\mathbf{S}$. In the equilibrium collinear state, the local moments are compensated, $\hat{\mathbf{m}}^\mathrm{A} = -\hat{\mathbf{m}}^\mathrm{B} = \hat{\mathbf{N}}$, where $\hat{\mathbf{N}}$ is the direction of the N\'eel vector. We have chosen the following parameters in our calculation, otherwise stated: $t_\pi= 0.5\, \text{eV}, \ \Delta = 1\, \text{eV}, \ J= 1\, \text{eV}, \ \lambda= 0.05\, \text{eV}$.


Figure~\ref{fig:band_structure}(b) shows the resulting electronic band structure of the model, where the spin expectation value projected in the N\'eel-vector direction is shown in color. This clearly shows the characteristic spin splitting of the $d$-wave altermagnet, which reverses the sign when $\mathbf{k}$ is rotated by $90^\circ$. As we illustrate in Fig.~\ref{fig:schematics}(c,d), the microscopic origin of the spin splitting lies in the orbital-dependent band splitting. Figures~\ref{fig:band_structure}(c) and \ref{fig:band_structure}(d) show the electronic band structures resolved by the OAM expectation values on sublattice A and B, respectively, along with the weights of the orbital characters that give rise to the corresponding OAM in each band. Near $E\approx -2.2~\mathrm{eV}$, where a band gap opens due to SOC, the OAM originates predominantly from sublattice A in this case. However, when the N\'eel vector is aligned along the $y$ axis instead of the $x$ axis, we find that sublattice B becomes the dominant contributor to the total OAM (see Supplemental Material \cite{supp} for details). This {\it alternating} nature of the sublattice-dependent OAM indicates that compensated magnetic order is intrinsically intertwined with a staggered orbital configuration arising from the mutually rotated electric quadrupolar order, as we have shown in Fig.~\ref{fig:schematics}(c,d). This coupling between spin order and the local orbital environment provides the microscopic foundation for the anisotropic orbital response discussed below.

To evaluate the OAM generated by antiferromagnetic resonance \cite{keffer1952theory}, we generalize the microscopic pumping formalism developed for ferromagnets~\cite{go2025orbital} to the case of two-sublattice altermagnet, whose detailed derivation can be found in \cite{supp}. Within the adiabatic approximation, the electronic states follow the slowly varying local moments $\hat{\mathbf m}^\mathrm{A(B)}(t)$ on site A(B). The pumping rate of the OAM is expressed as a response to the generalized forces, $\hat{\mathbf m}\times\partial_t\hat{\mathbf m}$,
\begin{eqnarray}
\partial_t\langle L_\alpha \rangle
&=&
\sum_{\beta=x,y,z}
\chi_{\alpha\beta}^\mathrm{A}
\left(
\hat{\mathbf m}^\mathrm{A}\times\partial_t\hat{\mathbf m}^\mathrm{A}
\right)_\beta
\nonumber 
\\ 
& & \ \ \ \ \ \ \ \  + 
\chi_{\alpha\beta}^\mathrm{B}
\left(
\hat{\mathbf m}^\mathrm{B}\times\partial_t\hat{\mathbf m}^\mathrm{B}
\right)_\beta
.
\label{eq:pumping_response}
\end{eqnarray}
The time-reversal-even contribution within the constant-broadening approximation, which eventually makes a dc signal, is given by
\begin{eqnarray}
& & 
\chi^{\mathrm{A(B)}}_{\alpha\beta}
=
\frac{\hbar}{\pi N_\mathbf{k}}
\sum_{\mathbf k}
\sum_{n,m}
A_{n\mathbf k}(E_F)
A_{m\mathbf k}(E_F)
\nonumber 
\\
& & \ \ \ \ \ \ \ \ \ \ \times
\operatorname{Re}
\left[
\langle n\mathbf k|
T^{L_{\alpha}}
|m\mathbf k\rangle
\langle m\mathbf k|
T^{S_{\beta}}_{\mathrm{A(B)}}
|n\mathbf k\rangle
\right].
\label{eq:pumping_coefficient}
\end{eqnarray}
Here,
\begin{gather}
T^{L_{\alpha}} 
= 
\frac{1}{i \hbar} \left[ L_{\alpha}, \mathcal{H}_{\mathrm{tot}} \right], 
\quad 
T^{S_{\beta}}_{\mathrm{A(B)}} 
= 
\frac{1}{i \hbar} \left[ S_{\beta}, \mathcal{H}^{\mathrm{A(B)}}_{\mathrm{XC}} \right] 
\end{gather}
are the orbital and spin torque operators, and 
\begin{equation}
\\
A_{n\mathbf k}(E_F)
=
\frac{\Gamma}
{(E_F-E_{n\mathbf k})^2+\Gamma^2}
\label{eq:spectral_function}
\end{equation}
is the spectral function with $\Gamma$ denoting the broadening constant. These formulas mean that the pumping rate of the OAM is determined by the torque-torque correlation between the spin and OAM, while the mixing of spectral weights between bands $n$ and $m$ is important. This therefore suggests crucial role of the orbital-dependent splittings in an altermagnet, which is alternating on different sites.

Figure~\ref{fig:band_structure}(e) shows the Fermi-energy dependence of the orbital pumping rate for $\mathbf{\hat{N}} \parallel \mathbf{\hat{x}}$ under right-handed (RH) magnon mode with cone angles $\theta_{A}=1.5^{\circ}$ and $\theta_{B}=1^{\circ}$, evaluated from Eq.~\eqref{eq:pumping_response}. For an energy broadening of $\Gamma = 25~\mathrm{meV}$, orbital pumping is enhanced in energy regions where the OAM is pronounced, including the previously mentioned hotspots near $E \approx -2.2~\mathrm{eV}$. Moreover, even when the energy broadening is increased to $\Gamma = 100~\mathrm{meV}$, an enhancement remains visible in the same energy region, although large $\Gamma$ suppresses the overall orbital-pumping response. This results indicates that the pumping is predominantly governed by the microscopic coherence between the $d_{xy}$ and $d_{zx}$($d_{yz}$) orbitals on sublattice A(B) for precession about $\mathbf{\hat{x}}$($\mathbf{\hat{y}}$).

While Fig.~\ref{fig:band_structure}(e) shows only the snapshot at $t=0$, Fig.~\ref{fig:band_structure}(f) shows the time evolution of the orbital pumping rate at $E_F=-2.2~\mathrm{eV}$ for the RH mode with $\hat{\mathbf{N}}\parallel\hat{\mathbf{x}}$ and for the LH mode with $\hat{\mathbf{N}}\parallel\hat{\mathbf{y}}$, shown in green solid line and purple dashed line, respectively. These two modes pump the OAM with exactly the same magnitude but opposite signs. Since the two sublattices of a $d$-wave altermagnet are exchanged by $\mathcal{C}_{4z}$, the two modes are related by $\mathcal{C}_{4z}\mathcal{T}$. We emphasize that this is a distinctive behavior that distinguishes an altermagnet from a conventional antiferromagnet.
Meanwhile, the orbital pumping rate exhibits a pronounced second-harmonic ($2\omega$) component, a characteristic feature of orbital pumping, as pointed out in Ref.~\cite{han2025orbital}. Our analysis of the spin response further reveals that, through SOC, spin pumping can also acquire a $2\omega$ component inherited from the orbital pumping, as also reported in Ref.~\cite{varela2023high} (see Supplemental Material \cite{supp} for details).

Integrating the pumping rate over a period gives the dc orbital
angular momentum accumulated per cycle,
\begin{equation}
\Delta \langle \mathbf{L} \rangle = \int_0^T dt\, \partial_t\langle \mathbf{L} \rangle.
\label{eq:pumped_angular_momentum}
\end{equation}
We evaluate this as the N\'eel vector is rotated within the $xy$ plane,
$\hat{\mathbf{N}}(\phi)=(\cos\phi,\sin\phi,0)$, and project the pumped OAM
onto $\hat{\mathbf{N}}$. Only even harmonics of $\phi$ survive, and our
model admits
\begin{equation}
\Delta \langle \mathbf{L} \cdot \hat{\mathbf{N}}(\phi) \rangle
= \Delta \langle L \rangle_{s}
+ \Delta \langle L \rangle_{d} \cos (2\phi)
+ \Delta \langle L \rangle_{g} \cos (4\phi).
\label{eq:decomposed_pumping}
\end{equation}
The two anisotropic terms have different origins. The $\cos4\phi$ term is
permitted by the $\mathcal{C}_{4z}$ symmetry of the square lattice and is
present for any magnetic order on this lattice, which is thus present also for conventional antiferromagnets. The $\cos2\phi$ term is
forbidden by that symmetry alone and becomes allowed only through the
sublattice-alternating crystal field. It is therefore specific to the
altermagnet.

\begin{figure}[t!]
\includegraphics[angle=0, width=0.47\textwidth]{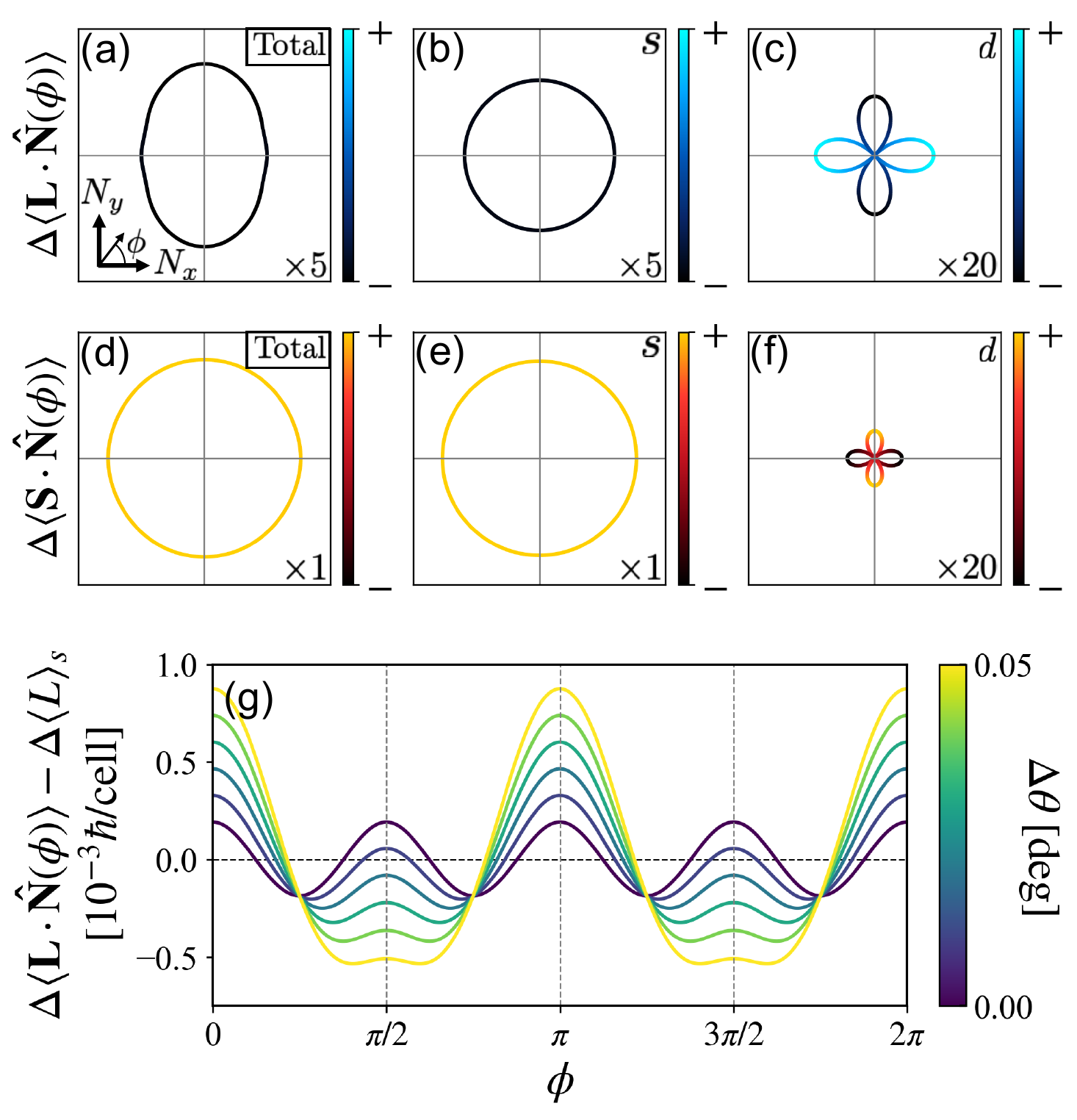}
\caption{
\label{fig:Dynamical_responses}
(a-f) Angular dependence of the dc orbital (a,b,c) and spin (d,e,f) pumping as N\'eel vector is rotated in the $xy$ plane at $E_{F}=-2.2~\mathrm{eV}$. (a,d) show the total pumping, while (b,e) and (c,f) show their $s$-wave and $d$-wave components, respectively. Note that while the spin pumping is dominated by the $s$-wave contribution, the orbital pumping exhibits sizable $d$-wave contribution, larger than that for the spin pumping. (g) Influence of the cone-angle difference of two sublattice magnetizations ($\Delta \theta \equiv \theta_{A} - \theta_{B}$) except for $s$-wave component. For $\Delta\theta=0$, the angular dependence of the orbital pumping exhibits $C_4$ symmetry ($g$-wave) due to cubic anisotropy of the crystal. However, as $\Delta\theta$ increases, $d$-wave component progressively becomes a more dominant contribution.
}
\end{figure}

Figure~\ref{fig:Dynamical_responses} compares the orbital and spin responses
at $E_F=-2.2~\mathrm{eV}$ for the RH mode, the spin pumping being obtained
by replacing $T^{L_\alpha}$ with $T^{S_\alpha}$ in
Eq.~\eqref{eq:pumping_coefficient}. Here, the magnitude is represented by the distance from the origin, and the sign is indicated by colors. The isotropic components behave
conventionally: spin pumping~[Fig.~\ref{fig:Dynamical_responses}(e)] dominates over the orbital pumping~[Fig.~\ref{fig:Dynamical_responses}(b)] by an order of magnitude, consistent with spin pumping as the primary response to
magnetization dynamics and orbital pumping as its SOC-mediated secondary
\cite{go2025orbital}. However, the anisotropic components reverse this ordering. The $d$-wave amplitude of orbital pumping~[Fig.~\ref{fig:Dynamical_responses}(c)] is larger than that for spin pumping~[Fig.~\ref{fig:Dynamical_responses}(f)] by factor 2.1 in this case (see End Matter for details). The asymmetry is already
visible in the torque operators of Eq.~\eqref{eq:pumping_coefficient}. The
orbital torque is a commutator with the full Hamiltonian and therefore
contains $\mathcal{H}_\mathrm{CF}$ explicitly, since $\mathbf{L}$ does not
commute with the crystal field. The spin torque is a commutator with the
exchange term alone and contains no crystal field at all. The staggered
order thus enters the orbital response at leading order and the spin
response only through SOC-induced mixing of the eigenstates.

In order to clarify the origin of the $d$-wave anisotropy, we examine how the angular dependence of the dc orbital pumping evolves as $\theta_{A}$ is varied while $\theta_{B}$ is fixed at $1^{\circ}$. Figure~\ref{fig:Dynamical_responses}(g) shows the anisotropic dc orbital pumping, decomposed into its $d$- and $g$-wave components, as the difference in the cone angles of the two sublattices ($\Delta\theta$) is progressively increased. Note that for visual clarify, we have subtracted the isotropic ($s$-wave) component. For $\Delta\theta = 0$, the angular dependence consists only of the $s$- and $g$-wave anisotropy, the latter originating from the intrinsic $\mathcal{C}_{4}$ crystal symmetry of the square lattice. As the cone-angle difference increases, however, the $d$-wave anisotropy is progressively enhanced (from purple to yellow colors). This behavior can be understood as follows: although each sublattice individually generates a $d$-wave orbital pumping arising from the altermagnetic crystal field in our model, the two contributions are opposite in sign and cancel exactly for identical cone angles. The cone-angle difference induced by antiferromagnetic resonance lifts this cancellation, leaving a finite $d$-wave orbital pumping. The emergence of the $d$-wave anisotropy therefore provides a direct fingerprint of the underlying $d$-wave local crystal field and the resulting electric quadrupolar order.

A concern for experiment is that thermal excitation may populate the RH and LH magnon modes equally, so that their pumping contributions cancel. Such a cancellation is indeed exact in a conventional antiferromagnet, where the two sublattices are related by the inversion $\mathcal{P}$, so that $\mathcal{P}\mathcal{T}$ symmetry forces the two modes to pump opposite amounts of OAM at the \emph{same} N\'eel orientation~\cite{cheng2014spin}. An altermagnet is defined by the absence of both operations: its sublattices are instead exchanged by $\mathcal{C}_{4z}$, so the modes are related by $\mathcal{C}_{4z}\mathcal{T}$ and are tied only at N\'eel directions differing by $90^\circ$,

\begin{equation}
\Delta\big\langle \mathbf{L}\cdot\hat{\mathbf{N}}\big\rangle_{\mathrm{LH}}(\phi)
= -\,\Delta\big\langle \mathbf{L}\cdot\hat{\mathbf{N}}\big\rangle_{\mathrm{RH}}\!\left(\phi-\tfrac{\pi}{2}\right).
\end{equation}
No operation maps the RH mode to the LH mode at fixed $\hat{\mathbf{N}}$, and the cancellation is therefore only partial. Expanding in harmonics as in Eq.~\eqref{eq:decomposed_pumping}, we obtain
\begin{equation}
\Delta\big\langle \mathbf{L}\cdot\hat{\mathbf{N}}\big\rangle_{\mathrm{RH}} + \Delta\big\langle \mathbf{L}\cdot\hat{\mathbf{N}}\big\rangle_{\mathrm{LH}} = 2 \Delta \langle L \rangle_{d}\cos (2\phi) + \cdots,
\end{equation}
in agreement with our numerical results~\cite{supp}. The thermally averaged signal is thus a pure $d$-wave response, making it an unambiguous fingerprint of the altermagnetic order. 


\begin{figure}[t!]
\includegraphics[angle=0, width=0.47\textwidth]{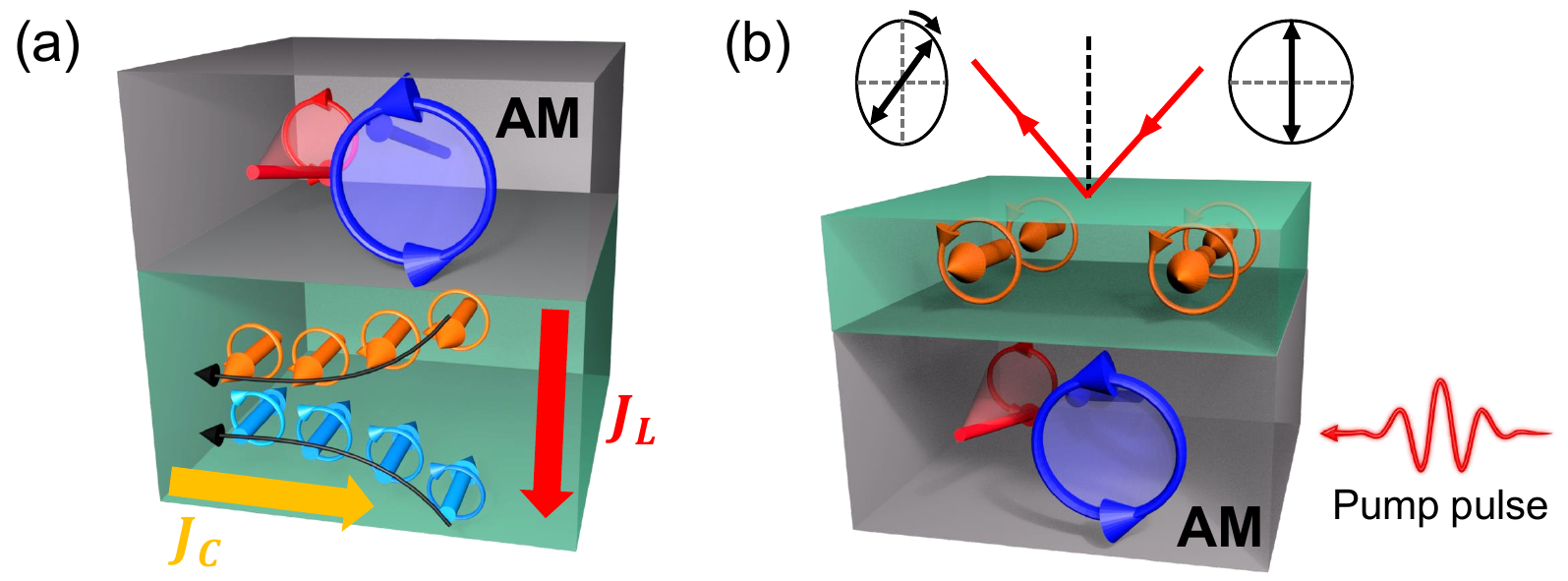}
\caption{
\label{fig:experimental_setups}
Schematic illustration of an experimental setup for detecting inherent electric quadrupole in $d$-wave AM via inverse orbital Hall effect (a) and magneto-optical Kerr effect (b), as N\'eel vector is rotated within the $xy$ plane.
}
\end{figure}

Finally, we propose two experimental schemes for detecting the anisotropic orbital pumping in altermagnets. In the first, the altermagnet is grown on top of a layer with a strong orbital Hall response, such as a light metal Ti, as illustrated in Fig.~\ref{fig:experimental_setups}(a). Under antiferromagnetic resonance, the pumped OAM flows into this underlayer as an orbital current, where the inverse orbital Hall effect converts it into a transverse charge current that is measured as a dc voltage~\cite{hayashi2024observation,wang2023inverse, choi2023observation}. Rotating the N\'{e}el vector within the $xy$ plane and recording the voltage traces out the angular dependence of Fig.~\ref{fig:Dynamical_responses}(g), from which the $d$- and $g$-wave components can be separated. A second and more direct route is a pump--probe measurement: a pump pulse excites the antiferromagnetic magnons and a delayed probe pulse detects the resulting OAM polarization by magneto-optical Kerr effect [Fig.~\ref{fig:experimental_setups}(b)]. Repeating this for different N\'{e}el-vector orientations reveals the same anisotropy without the need for a conversion layer.

In summary, we have established anisotropic orbital pumping driven by antiferromagnetic magnons as a macroscopic fingerprint of $d$-wave altermagnet. Using a microscopic pumping formalism, we showed that antiferromagnetic resonance generates both spin and orbital angular momentum in itinerant electrons, with the orbital pumping exhibiting a substantially stronger anisotropy than the spin pumping. This enhanced anisotropic orbital pumping originates from the direct coupling of the orbital response to the staggered local crystal field, whereas the corresponding spin response probes it only indirectly through SOC. We further demonstrated that the angular dependence contains a pronounced $d$-wave component associated with the staggered altermagnetic crystal-field order, distinct from the contribution imposed by the underlying lattice symmetry. These results establish a direct connection among antiferromagnetic magnons, altermagnetic crystal-field order, and orbitronics, demonstrating that orbital pumping provides a dynamical fingerprint of $d$-wave altermagnet and opening new opportunities for magnon-driven orbitronics in the terahertz regime.

\begin{acknowledgements}
This work was supported by a Korea University Grant, the POSCO Science Fellowship of POSCO TJ Park Foundation, and the National Research Foundation of Korea (NRF) grant funded by the Korea government (MSIT) (No. RS-2024-00410027, RS-2026-25468443). K.-W. K was financially supported by the NRF (No. RS-2024-00334933, RS-2026-25543309) H.-W. L was financially supported by the NRF (No. RS-2024-00356270, RS-2026-25585727).
\end{acknowledgements}

\bibliography{reference}


\section*{END MATTER}
\begin{center}
    \textbf{Quantitative Comparison among a Ferromagnet, an Antiferromagnet, and an Altermagnet}
\end{center}




We quantitatively compare the isotropic ($s$-wave) and anisotropic ($d$- and $g$-wave) components of dc orbital pumping in Ferromagnet (FM), Antiferromagnet (AF), and $d$-wave AM at the $E_{F}=-2.2~\mathrm{eV}$ for $\Gamma=25~\mathrm{meV}$ and $100~\mathrm{meV}$ in Table~\ref{tab:comparision1}. Among the three magnetic phases, the $d$-wave AM exhibits the largest anisotropy relative to its isotropy in orbital-pumping response. This pronounced anisotropy is nearly unchanged as $\Gamma$ increases, demonstrating that it is robust with respect to the degree of disorder. Furthermore, our result that coherent resonance in antiferromagnets gives rise to finite orbital pumping, which is consistent with the results of Cheng \textit{et al}.~\cite{cheng2014spin} and Vaidya \textit{et al}.~\cite{vaidya2020subterahertz}, who showed that spin-pumping contributions from antiparallel sublattices constructively reinforce each other rather than cancellation.

\begin{table}[h]
    \centering
    \caption{
        \label{tab:comparision1}
        $\Delta \langle L \rangle_{s}$, $\Delta \langle L \rangle_{d} + \Delta \langle L\rangle_{g}$ and their ratio ($(\Delta \langle L \rangle_{d} + \Delta \langle L\rangle_{g}) / \Delta \langle L \rangle_{s}$) of orbital pumping for FM, AF, and $d$-wave AM. Values without (within) parentheses correspond to $\Gamma=25~\mathrm{meV}$ ($100~\mathrm{meV}$). All values are given in units of $10^{-3}\hbar/\mathrm{cell}$ except for ratio.
    }
    \begin{ruledtabular}
        \begin{tabular}{c|ccc}
            Material\rule[-1.7ex]{0pt}{3.8ex}
            & \makecell{$\Delta \langle L \rangle_{s}$}
            & \makecell{$\Delta \langle L \rangle_{d} + \Delta \langle L \rangle_{g}$}
            & \makecell{Ratio [\%]}
            \\
            \hline
            \noalign{\vskip 4pt}
            FM
            & \makecell{$1.09\times 10^{-1}$ \\ ($2.07\times 10^{-1}$)}
            & \makecell{$1.14\times10^{-3}$ \\ ($1.35\times 10^{-3}$)}
            & \makecell{$1.04$ \\($6.51\times 10^{-1}$)}
            \\[10pt]
            AF
            & \makecell{$3.72$ \\ ($6.75\times 10^{-1}$)}
            & \makecell{$8.69\times 10^{-6}$ \\ ($1.28\times 10^{-6}$)}
            & \makecell{$2.33\times 10^{-4}$ \\ ($1.89\times 10^{-4}$)}
            \\[10pt]
            AM
            & \makecell{$4.39$ \\ ($1.01$)}
            & \makecell{$9.66\times 10^{-1}$ \\ ($5.81\times 10^{-1}$)}
            & \makecell{$22$ \\ ($57.5$)}
        \end{tabular}
    \end{ruledtabular}
\end{table}

\begin{center}
    \textbf{Spin versus Orbital Pumping in an Altermagnet}
\end{center}

As shown in Fig.~\ref{fig:Dynamical_responses}, the $d$-wave component of orbital pumping is larger than that of spin pumping, whereas its $s$-wave component is smaller. Here, we quantitatively compare these angular dependences by decomposing the spin- and orbital-pumping responses into their $s$- and $d$-wave components. Table~\ref{tab:comparision2} summarizes the each component and their ratio, $\Delta\langle L\rangle_d/\Delta\langle L\rangle_s$ at $E_F=-2.2~\mathrm{eV}$. The ratio is substantially larger for orbital pumping than for spin pumping, quantitatively supporting orbital pumping as a dynamical fingerprint of $d$-wave AM.

\begin{table}[h]
    \centering
    \caption{
        \label{tab:comparision2}
        $\Delta\langle L\rangle_s$, $\Delta\langle L\rangle_d$, and their ratio, $\Delta\langle L\rangle_d/\Delta\langle L\rangle_s$, for spin and orbital pumping in a $d$-wave AM at $E_F=-2.2~\mathrm{eV}$. Values outside (inside) parentheses correspond to $\Gamma=25~\mathrm{meV}$ ($100~\mathrm{meV}$). All values are given in units of $10^{-3}\hbar/\mathrm{cell}$.
    }
    \begin{ruledtabular}
        \begin{tabular}{c|ccc}
            Pumping\rule[-1.7ex]{0pt}{3.8ex}
            & \makecell{$\Delta \langle L \rangle_{s}$}
            & \makecell{$\Delta \langle L \rangle_{d}$}
            & \makecell{$\Delta \langle L \rangle_{d} / \Delta \langle L \rangle_{s}$ [\%]}
            \\
            \hline
            \noalign{\vskip 4pt}
            Orbital
            & \makecell{$4.39$ \\ ($1.01$)}
            & \makecell{$0.86$ \\ ($0.57$)}
            & \makecell{$19.66$ \\($56.52$)}
            \\[10pt]
            Spin
            & \makecell{$28.42$ \\ ($26.95$)}
            & \makecell{$0.40$ \\ ($0.11$)}
            & \makecell{$1.41$ \\ ($0.41$)}
        \end{tabular}
    \end{ruledtabular}
\end{table}

\begin{center}
    \textbf{Two-Ion Cluster Model}
\end{center}

To elucidate the role of the local crystal field in the anisotropy of orbital pumping, independently of its kinetic contribution, we construct a single-ion model for sublattice A in Fig.~\ref{Fig1}(a). In the Hilbert space spanned by $t_{2g}$ orbital basis, the Hamiltonian is given by
\begin{eqnarray}
\mathcal{H} 
= 
\tilde{J} \mathbf{\hat{m}} \cdot \mathbf{L} 
+ 
\Delta \mathrm{diag}[-1,\ +1,\ -1]
\end{eqnarray}

\noindent with orbital angular momentum operators

\begin{gather}
    L_{x} = \begin{pmatrix}
        0 & 0 & -i \\
        0 & 0 & 0 \\
        i & 0 & 0
    \end{pmatrix}, \quad 
    L_{y} = \begin{pmatrix}
        0 & i & 0 \\
        -i & 0 & 0 \\
        0 & 0 & 0
    \end{pmatrix}, \notag \\
    L_{z} = \begin{pmatrix}
        0 & 0 & 0 \\
        0 & 0 & i \\
        0 & -i & 0
    \end{pmatrix}
\end{gather}

The first term describes the effective exchange coupling between the magnetization $\mathbf{\hat{m}}$ and the orbital angular momentum $\mathbf{L}$, with $\tilde{J}$ incorporating the effect of SOC, while the second term describes the anisotropic crystal-field potential.

\begin{figure}[b!]
\includegraphics[angle=0, width=0.47\textwidth]{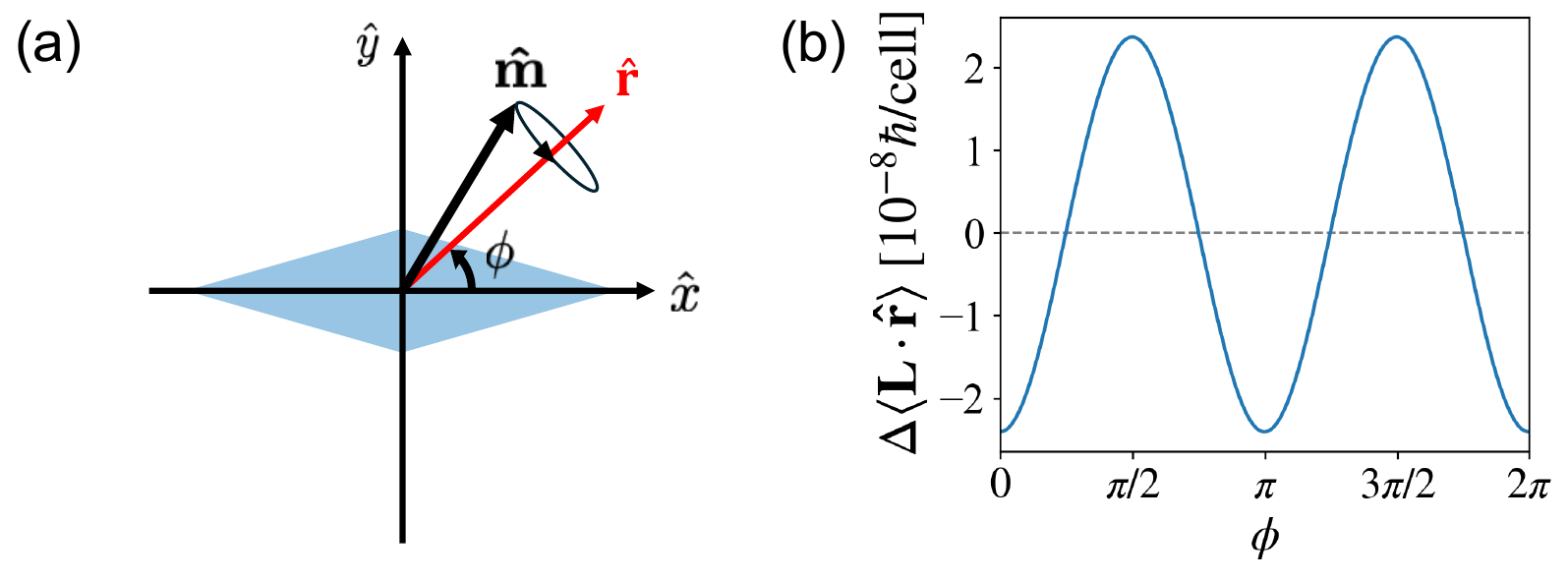}
\caption{
\label{fig:single_ion}
(a) Schematic illustration for single-ion model with precession axis $\mathbf{\hat{r}}$ and magnetization $\mathbf{\hat{m}}$. (b) Angular dependence of orbital pumping from single-ion model.
}
\end{figure}

We consider a configuration in which the magnetization $\mathbf{\hat{m}}$ precesses about the axis $\mathbf{\hat{r}} = (\cos{\phi}, \sin{\phi}, 0)$ with a cone angle $\theta_{c}$ illustrated in Fig.~\ref{fig:single_ion}(a). Retaining terms up to first order in $\theta_{c}$, the magnetization is expressed as

\begin{equation}
    \mathbf{\hat{m}} = \begin{pmatrix}
        \cos{\phi} - \theta_{c} \sin{\phi} \sin{\omega t} \\
        \sin{\phi} + \theta_{c} \cos{\phi} \sin{\omega t} \\
        -\theta_{c} \cos{\omega t}
    \end{pmatrix}.
    \label{eq:single_magnetization}
\end{equation}

To apply degenerate perturbation theory in the limit $\tilde{J} \ll \Delta$, we recast the Hamiltonian in the form

\begin{equation}
    \mathcal{H} = \hat{H}_{0} + \tilde{J} \hat{V},
\end{equation}

\noindent where

\begin{equation}
    \hat{H}_{0} = \Delta \mathrm{diag}[-1,\ +1,\ -1], \quad \hat{V} = \hat{m} \cdot \mathbf{L}.
\end{equation}

The wavefunctions of the unperturbed Hamiltonian $\hat{H}_{0}$ and their corresponding eigenenergies are

\begin{equation}
    |\psi_{1}^{(0)} \rangle = \frac{1}{\sqrt{2}} \begin{pmatrix}
        1 \\
        0 \\
        i
    \end{pmatrix}, ~
    |\psi_{2}^{(0)} \rangle = \begin{pmatrix}
        0 \\
        1 \\
        0
    \end{pmatrix}, ~
    |\psi_{3}^{(0)} \rangle = \frac{1}{\sqrt{2}} \begin{pmatrix}
        -1 \\
        0 \\
        i
    \end{pmatrix}
\end{equation}

\noindent with

\begin{equation}
    E_{1}^{(0)} = E_{3}^{(0)} = -\Delta,\quad E_{2}^{(0)} = \Delta.
\end{equation}

For notational convenience, we introduce $a\equiv m_{x}$, $b\equiv m_{y}$, and $c\equiv m_{z}$ in Eq.~\eqref{eq:single_magnetization}. Applying first-order perturbation theory, we obtain perturbed wavefunctions and their energies, are given by

\begin{align}
    |\psi_{1} \rangle &= \left[ \hat{I}_{3} - \frac{\tilde{J}}{2 \Delta} \left( \hat{V} - a \hat{I}_{3} \right) \right] | \psi_{1}^{(0)} \rangle, && E_{1} = -\Delta + \tilde{J} a \notag \\
    |\psi_{2} \rangle &= \left[ \hat{I}_{3} + \frac{\tilde{J}}{2 \Delta} \hat{V} \right] | \psi_{2}^{(0)} \rangle, && E_{2} = \Delta \notag \\
    |\psi_{3} \rangle &= \left[ \hat{I}_{3} - \frac{\tilde{J}}{2 \Delta} \left( \hat{V} + a \hat{I}_{3} \right) \right] | \psi_{3}^{(0)} \rangle, && E_{3} = -\Delta - \tilde{J} a,
\end{align}

\noindent where $\hat{I}_{3}$ is $3\times 3$ identity matrix.

Substituting these states into Eq.~\eqref{eq:pumping_response} and neglecting all terms of order $\mathcal{O}(\tilde{J}/\Delta)$, we obtain

\begin{equation}
    \Delta \langle \mathbf{L}\cdot \mathbf{\hat{r}} \rangle_{\mathrm{single}} (\phi) = -\tilde{J}^{2} \theta_{c}^{2} \omega \left[ C_{s} + C_{d} \cos{(2\phi)} \right] \mathcal{A}_{\mathrm{CF}} (E_{F}),
    \label{eq:single_ion}
\end{equation}

\noindent where the coefficients are defined as

\begin{gather}
    C_{s} \equiv \frac{2}{T\Gamma} - \frac{1}{T^{2} \Gamma} \left( \frac{\tilde{J}}{\Gamma} \right)^{2} \qquad C_{d} \equiv -\frac{1}{T^{2} \Gamma} \left( \frac{\tilde{J}}{\Gamma} \right)
\end{gather}

\noindent with $T \equiv \left[ (E_{F} + \Delta)/ \Gamma \right]^{2} + 1$, and the spectral function is defined as

\begin{equation}
    \mathcal{A}_{\mathrm{CF}} (E_{F}) \equiv \frac{\Gamma}{(E_{F} - \Delta)^{2} + \Gamma^{2}}.
\end{equation}


In conclusion, equation~\ref{eq:single_ion} shows that the local crystal field gives rise to a $d$-wave anisotropy in orbital pumping. The angular dependence of orbital pumping dominantly arises from perturbed eigenenergies induced by the local crystal field in first-order perturbative regime with strong crystal-field potential, while in the exact-diagonalization calculation, coherence also contributes to the angular dependence as shown in Fig.~\ref{fig:single_ion}(b). In addition, the orbital pumping from the other sublattice can be obtained by replacing $\phi$ with $\phi+\pi/2$. Combining the orbital pumping from the two sublattices demonstrates that the difference in the cone angles induced by antiferromagnetic resonance prevents cancelation of each $d$-wave anisotropic orbital pumping, thereby revealing the intrinsic $d$-wave anisotropy of the AM, consistent with the results in Fig.~\ref{fig:Dynamical_responses}(g).



\end{document}